\documentclass[]{article}
\usepackage{authblk}
\usepackage{epsfig}
\usepackage{graphicx}
\usepackage{subfigure}

\title{Influence of the particle shape on the equilibrium morphologies of supracolloidal magnetic filaments}

\author[1]{E.S. Pyanzina}
\author[1]{E.V. Novak}
\author[2]{D.A. Rozhkov}
\author[1]{A.V. Gudkova} 
\author[2]{P.A. S\'anchez}

\affil[1]{Ural Federal University, Leviv av. 51, 620000, Ekaterinburg}
\affil[2]{University of Vienna, Sensengasse 8, 1090, Vienna, Austria} 

\begin{document}
\maketitle


\begin{abstract}
We investigate the equilibrium morphologies of linear and ring-shaped magnetic filaments made from crosslinked ferromagnetic spherical or ellipsoidal colloidal particles. Using Langevin dynamics simulations, we calculate the radius of gyration and total magnetic moment of a single filament at zero field and different temperatures, analyzing the influence of the particles shape, the strength of their magnetic moment and the filament length. Our results show that, among such parameters, the shape of the particles has the strongest qualitative impact on the equilibrium behavior of the filaments.
\end{abstract}


\section{Introduction.}
Magnetic colloidal particles and materials based on them have been studied for already several decades. Conventional magnetic fluids, gels and elastomers are usually produced by means of relatively simple techniques. This provides a limited control of their microstructure and magnetic response. A novel and promising approach is the creation of small aggregates of magnetic particles with well defined shapes, permanently stabilized by polymer crosslinking in order be operated as building blocks of more sophisticated magnetoresponsive materials. Using modern experimental techniques like DNA crosslinking \cite{2014-byrom} or DNA designed self-assembly \cite{2014-srivastava}, one can have the possibility to create rather flexible linear polymer-like chains of magnetic colloidal particles, also known as magnetic filaments. On the other hand, in recent years, magnetic anisotropic particles have become an independent and fast growing branch in dipolar soft matter research. The study of anisotropic magnetic particles addresses particles with internal anisotropies like capped colloids and particles with shifted dipoles \cite{sacanna12a} or Janus particles \cite{Perro05a} and particles with shape anisotropies like magnetic rods or ellipsoids \cite{sacanna06a}. In this paper, for the first time we study magnetic filaments made of ellipsoidal magnetic particles by comparing their equilibrium behavior with the corresponding to filaments of spherical particles.

\section{Magnetic filament model.}


To represent magnetic filaments, we employ a bead-spring model introduced in our previous works \cite{2015-sanchez-mm1}. The magnetic beads carry a point magnetic dipole of moment $\vec{m}$ at their centers of geometry. The magnetic interactions are, therefore, calculated by means of the dipole-dipole pair potential:
\begin{equation}\label{eq-dip-dip}
U_d (ij) =- \left[ 3 \frac{\left( {{\vec m_i}} \cdot {\vec r_{ij}} \right) \left({\vec m_j}\cdot {\vec r_{ij}}\right)}{r_{ij}^5} - \frac{\left({\vec m_i}\cdot {\vec m_j} \right)}{r_{ij}^3} \right],
\end{equation}
where ${\vec r}_{ij}$ is the displacement vector connecting the centers of geometry of the particles $i$ and $j$. We model the particle's shape anisotropy using a orientation-dependent steric interaction. This is a modified Gay-Berne potential \cite{gay81a} that allows to represent ellipsoids with any arbitrary aspect ratio, $X_0 = a/b$, where $a$ and $b$ are the main geometry axes of the body, so that $X_0=1$ for the sphere and $X_0>1$ for the ellipsoid. This potential has the form:
\begin{equation}\label{eq-GB-general}
U_{GB}(ij)=\left \{  \begin{array}{c}
4 \varepsilon({\vec u}_i,{\vec u}_j)\left [ \left( \frac{\sigma_0}{r_{ij}-\sigma(\cdot)+\sigma_0}  \right)^{12}-\left( \frac{\sigma_0}{r_{ij}-\sigma(\cdot)+\sigma_0}  \right)^{6} + \frac{1}{4}\right ] , r_{ij}\leq r_c
 \\ 0,  r_{ij} > r_c.
\end{array}
 \right.
\end{equation}
where ${\vec u}_{i(j)} $ is the unit vector along the axis $b$ of the particle $i(j)$, $\sigma_0=2 \sqrt 2 a$, $\sigma(\cdot)=\sigma({\vec u}_i,{\vec u}_j,\hat{{ r}}_{ij})$ is the effective interparticle distance and $\varepsilon_0$ denotes the energy scale of the interaction. Finally, the permanent crosslinks between particles are represented as a harmonic potential that connects neighbor particles along the chain. In the case of spherical particles, this bonding potential connects their surfaces \cite{2015-sanchez-mm1}, whereas for ellipsoids, it connects their centers of geometry. In spheres, the dipole moment remains oriented in the direction defined by the surface crosslinks, whereas in ellipsoids is pointing parallel to their axis $a$. With this model, we performed Langevin dynamics simulations in the NVT ensemble by means of the ESPResSo simulation package \cite{arnold04b}. More details of the simulation protocol and parameters are presented in \cite{2015-sanchez-mm1, 2017-rozhkov}.


\section{Results and discussion.}


Here, we express all the physical parameters in a typical system of reduced units \cite{2017-rozhkov}.

\begin{figure}[!h]
\begin{center}
\includegraphics[width=11cm]{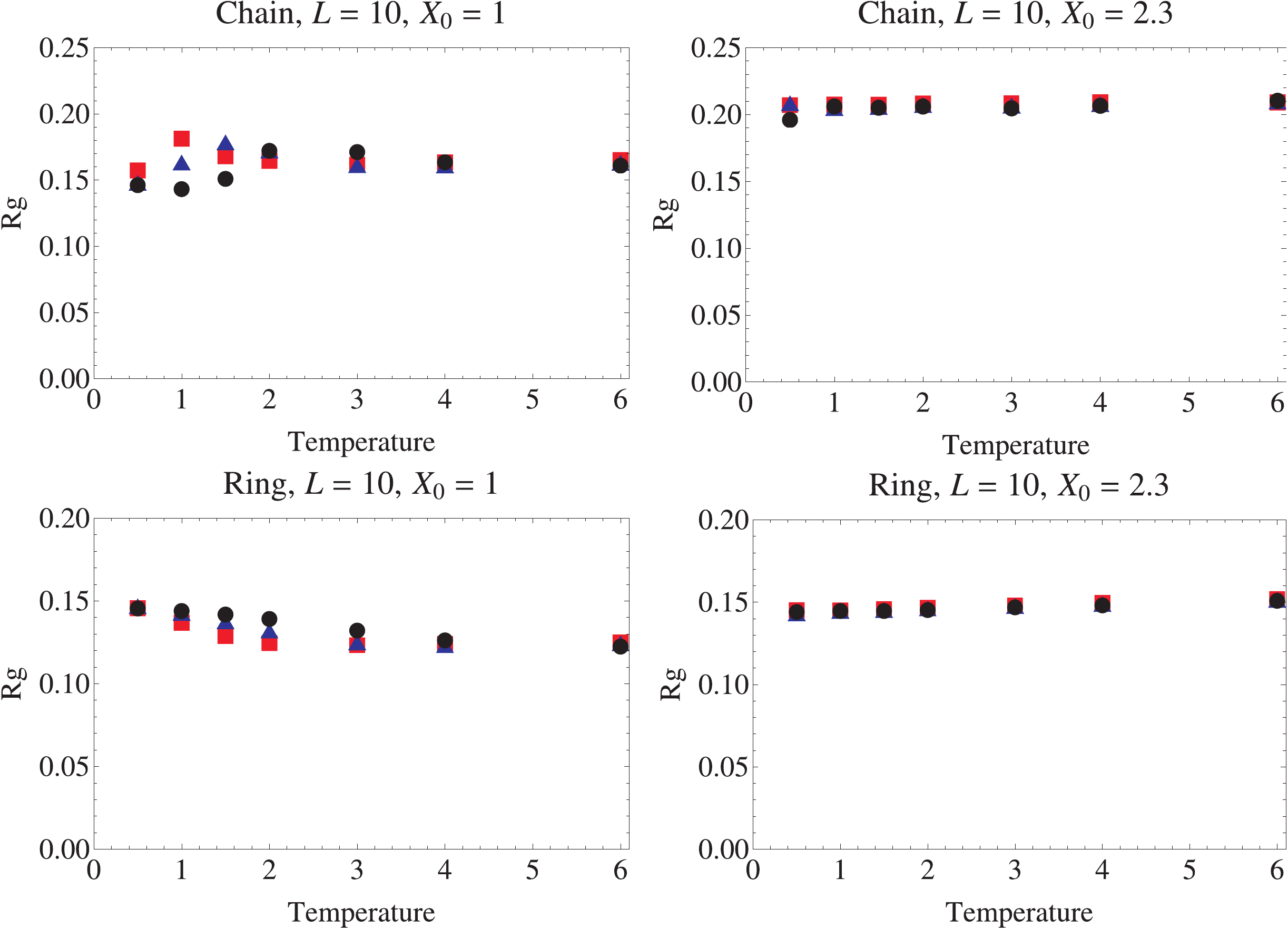}
\caption{The radius of the gyration as a function of the temperature. First column: results for filament composed of spheres ($X_0=1$), second column: a filament composed of ellipsoids (($X_0=2.3$)). Different symbols correspond to different values of the magnetic moment: squares - $m^2=3$, triangles - $m^2=5$, dots - $m^2=8$. }
\label{Fig:Rg}
\end{center}
\end{figure}

\begin{figure}[!h]
\begin{center}
\includegraphics[width=10cm]{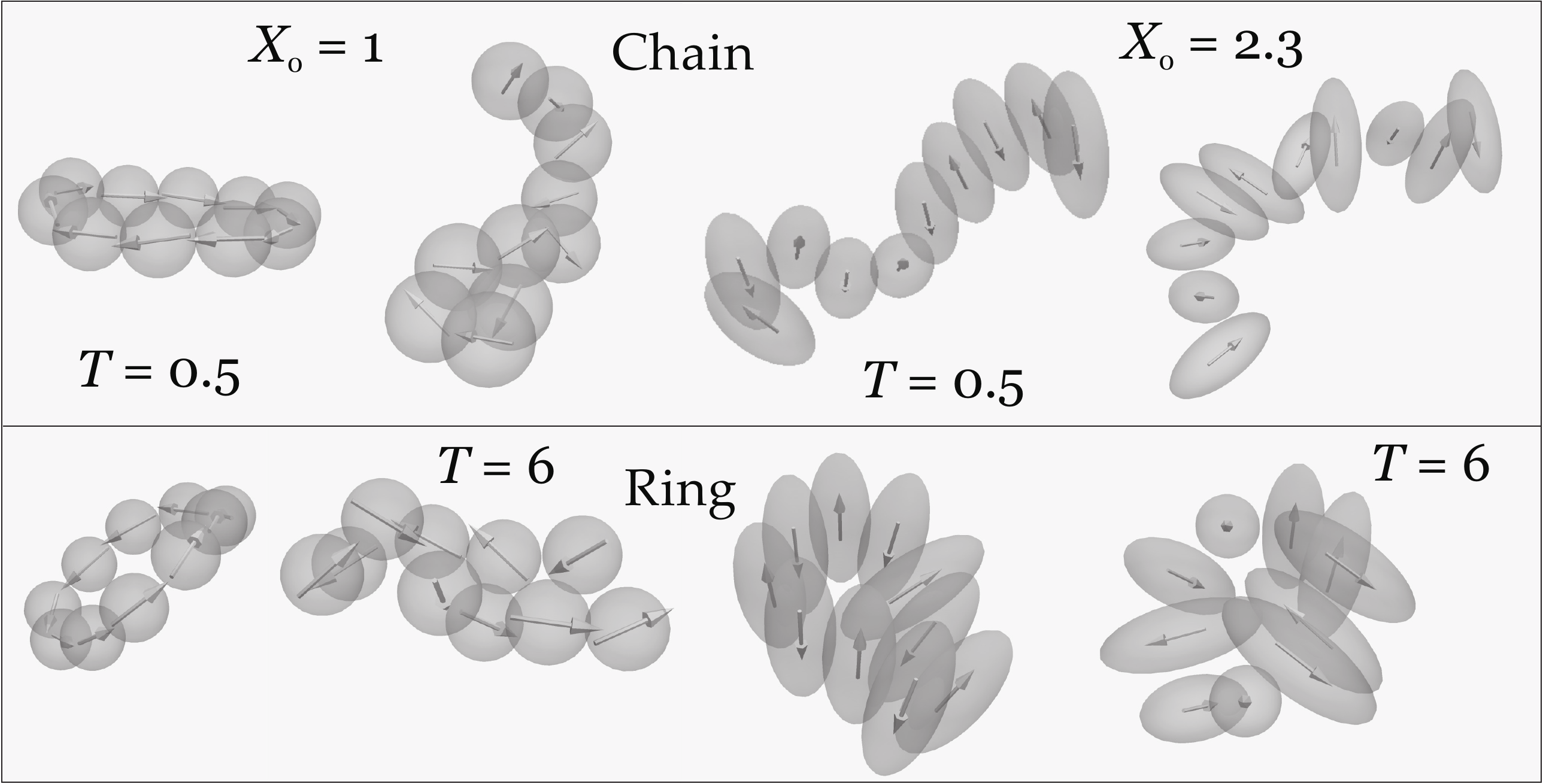}
\caption{Some typical configuration snapshots for different system parameters.}
\label{Fig:Snapshots}
\end{center}
\end{figure}

The first parameter which was calculated for the characterization of the equilibirum structures of the filaments is the radius of gyration defined in terms of the filament length $L$, the position of the center of mass of the filament $\vec R_{cm}$ and the positions of each one of its particles $\vec r_i$ as
\begin{equation}
R_g=\left [ \frac{1}{L}\sum_{i=1}^L{\left(\vec{r}_i-\vec{R}_{cm}\right)}^2 \right ]^{1/2}
.\end{equation}
Fig. \ref{Fig:Rg} shows the results of $R_g$ as a function of the temperature for the case $L=10$. For open filaments with spherical particles, we can see that $R_g$ grows with decreasing temperature until it experiences a drop, corresponding to the closure of the open chain into a ring (see snapshots in Figure \ref{Fig:Snapshots}). As known, the temperature at which the closure takes place depends on both the magnetic moment of the particles and the filament length. Filaments with ellipsoids do not experience any closure and their $R_g$ remains nearly constant within the sampled range of temperatures. This is due to the fact that ellipsoids can only form rings by connecting the filament free ends with their dipoles forming an antiparallel pair (see \cite{2013-kantorovich-sm}), which is energetically much less favorable than the head-to-tail pair formed in the closure of spherical particles. The same qualitative difference is observed for rings: while $R_g$ slightly decreases with temperature for the case of spherical particles, the rings of ellipsoids, in turn, keep an almost constant value. This proves that the rings of spheres are more flexible than rings of ellipsoids due to the geometrical constraints of the latter.

\begin{figure}[!h]
\begin{center}
\includegraphics[width=11cm]{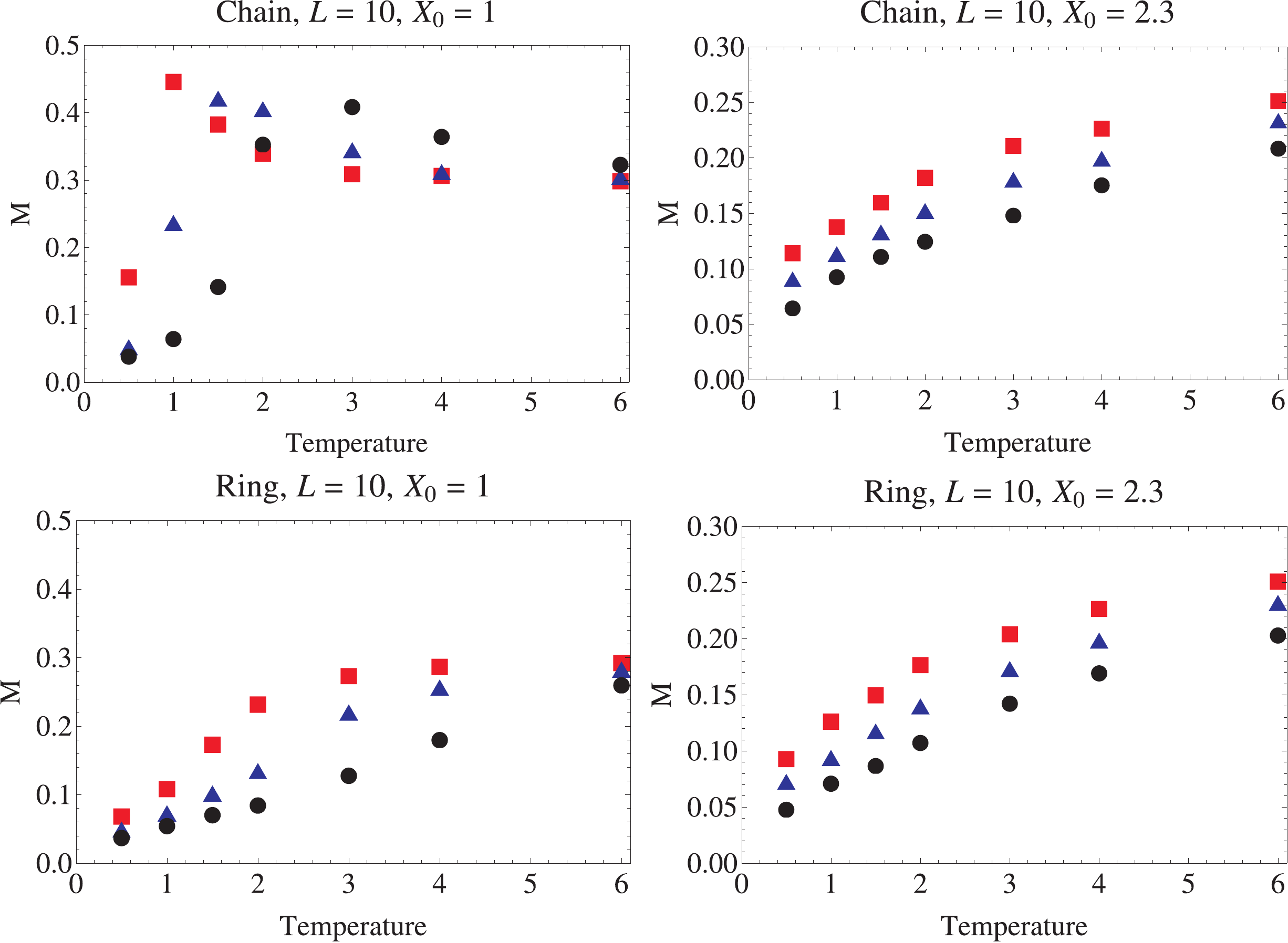}
\caption{Total magnetic moment of filament as function of temperature. The same arrangement as in Figure \ref{Fig:Rg}.}
\label{Fig:M}
\end{center}
\end{figure}

The second parameter analyzed is the relative net magnetic moment of the filament $M$ given by the normalized vector sum of the dipolar moments of its particles:
\begin{equation}
M=\frac{1}{L m}\left|\sum_{i=1}^L\vec{m}_i\right|.
\label{eq:magmom}
\end{equation}
Fig. \ref{Fig:M} displays the results for $M$ obtained under the same conditions as discussed above. Here, the closure of the open filaments with spheres is much more clearly signaled as $M$ drops to almost zero for closed filaments. The rest of cases, \textit{i.e.} open filaments with ellipsoids and both types of rings, show a monotonous decrease of $M$ at cooling, but the filaments of ellipsoids have values of $M$ significantly lower than their spherical counterparts. This is due to the alternating antiparallel orientation of the dipoles which they tend to keep independently on their open or closed arrangement, as the snapshots in Fig. \ref{Fig:Snapshots} illustrate.

Finally, results for longer filament lengths ($L=\lbrace 15,\,30 \rbrace$, not shown) exhibit the same qualitative behavior, with only moderate quantitative differences.



\section{Conclusions.}
We studied for the first time the equilibrium behavior of magnetic filaments made of ellipsoidal particles by means of computer simulations. Our results give evidence that the anisotropy of the particles leads to significant qualitative changes in equilibrium configurations within the range of explored parameters, replacing the structural closure observed for filaments with spherical particles at cooling by monotonous behaviors. We expect these results to be the basis of the future theoretical models and to be valuable for the design of novel magnetic systems.
\section*{Acknowledgements.}

This research was supported by the Ministry of Education and Science of Russia (Project 3.1438.2017/4.6). P.A.S. acknowledges support of the Austrian Science Fund (FWF):START-Projekt Y 627-N27.


\clearpage


\end{document}